# Science Learning via Participation in Online Citizen Science

**Karen Masters, Eun Young Oh, Joe Cox, Brooke Simmons, Chris Lintott, Gary Graham, Anita Greenhill, Kate Holmes**

*ABSTRACT: We investigate the development of scientific content knowledge of volunteers participating in online citizen science projects in the Zooniverse (www.zooniverse.org). We use econometric methods to test how measures of project participation relate to success in a science quiz, controlling for factors known to correlate with scientific knowledge. Citizen scientists believe they are learning about both the content and processes of science through their participation. We don't directly test the latter, but we find evidence to support the former - that more actively engaged participants perform better in a project-specific science knowledge quiz, even after controlling for their general science knowledge. We interpret this as evidence of learning of science content inspired by participation in online citizen science.*

**Keywords:** Citizen Science, Public engagement with science and technology, Public understanding of science and technology.

**Context**

Citizen science is defined as "scientific work undertaken by members of the general public, often in collaboration with or under the direction of professional scientists and scientific institutions" (Martin 2014). It is not a new concept, even if the definition only entered the dictionary recently. Notable early examples of citizen science include a call by astronomer Edmund Halley for observations of a total eclipse of the Sun that crossed central England in 1715 (Halley 1714), and in the late 19$^{th}$/early 20$^{th}$ century, the Audubon Society's Christmas Bird Count (Root 1988) and the American Association of Variable Star Observers (Williams 2001). These projects are normally remembered for their contribution to the advancement of science, but as large-scale public participation events they are considered to have a significant role to play in science education as well.

It is widely considered that participation in citizen science has the potential to lead to increased scientific literacy (by which we mean an increased understanding of both the content of science and the scientific process, as well as the contexts through which science occurs, e.g. Miller 2001, Lang et al. 2006, Bauer et al. 2007), primarily via the resulting exposure to authentic scientific practices (Brossard et al. 2005, Bonney et al. 2009, Raddick et al. 2010, Kloetzer 2013, Curtis 2015). Engaging in citizen science allows people to experience first-hand the scientific process and engage scientific thinking at the same time as increasing their knowledge of the specific research topic (i.e. their knowledge of scientific content). For example in order to participate in Halley's 1715 experiment, citizens had to learn about solar eclipses as well as the process by which to record the times of the eclipse.

Large scale citizen science projects predate the widespread adoption of the internet, and many published studies on the development of scientific literacy through involvement in citizen science focus on involvement of volunteers in what are primarily offline citizen science projects (e.g. Trumbull et al. 2000, Brossard et al. 2005, Evans 2005, Cronje 2011). To participate in offline projects volunteers must invest enough time to go out of doors and often collect relatively complex data (e.g. the volunteers studied in both Evans 2005 and Brossard et al. 2005 had to install nest boxes and capture long term data about visiting birds). These citizen scientists may be asked to devote considerable effort to the project (e.g. the 45 citizen scientists surveyed in Cronje 2011 were tested after participating in a 2-day event). While these studies have struggled to demonstrate any significant increase in scientific literacy in general, they have typically found evidence that participants increase their scientific knowledge about the topic of project (Brossard et al. 2005, Cronje 2011). Furthermore Trumbull et al. (2000) was able to qualitatively demonstrate the use scientific thinking during the engagement, and Evans et al. (2005) argued that both development of science content knowledge and scientific thinking were demonstrated by participants.

A host of platforms for online citizen science now allow anyone with access to the Internet to become a citizen scientist with the investment of much less time than offline (data collection) projects. Volunteers participate in online citizen science for a variety of reasons (Raddick et al. 2010), amongst them are the desire to learn more about a subject. Furthermore, science content learning can be shown to take place amongst at least a subset of volunteers in many online projects (e.g. Prather et al. 2013; Luczak-Rosch et al 2014). However, these two goals - of education and of scientific productivity - may often be in conflict, with time spent by project organizers furthering one not being spent on furthering the other. Particularly in modern, distributed data analysis tasks (Simpson 2014), the focus on useful 'work' dictated by scientific urgency may prevent the explicit design of such projects for the encouragement of learning about the content of the science topic (e.g. the set of images needing classifying to further science, may not be the best set to teach beginners subject knowledge), even while involvement in the project teaches by experience about the scientific method.

*Introduction to The Zooniverse*

The largest of all online citizen science platforms is the Zooniverse (www.zooniverse.org), run by the Citizen Science Alliance (CSA; www.citizensciencealiance.org). The Zooniverse currently has more than 1.4 million registered users, and hosts a selection of ~40 active online citizen science projects, where volunteers analyse data needed for academic research. Zooniverse projects cover areas as diverse as astronomy, climatology, genetics, papyrology and modern history. Participants in the Zooniverse can even assist in the study the history of citizen science itself[1].

[1] "Science Gossip": www.sciencegossip.org.

The first Zooniverse project was Galaxy Zoo (www.galaxyzoo.org). The phenomenal success of Galaxy Zoo after it launched in July 2007 (e.g. the project received 8 million classifications in its first 10 days) provided the inspiration for the creation of the Zooniverse in 2010. Galaxy Zoo shows volunteers images of galaxies (at first from the Sloan Digital Sky Survey Legacy Survey (York et al. 2000); more recently from other surveys including public Hubble Space Telescope Surveys) and asks a series of questions in a "decision tree" which lead the volunteer to describe common galactic morphologies. Galaxy Zoo publications have demonstrated the accuracy of a collective of human eyes in performing this task compared to relatively simple computer algorithms (e.g. Lintott et al. 2011, Willett et al. 2013).

Zooniverse projects have a common philosophy of making use of the citizen scientist input for peer reviewed academic research. To date, at least thirteen Zooniverse projects have resulted in a total of over 90 published results[2]. To qualify as a Zooniverse project the research team must have a task that is impossible (or very difficult) for computers to perform, more data than is practical for a small number of people to analyse and a genuine research question/need. Science education is considered in the development of Zooniverse projects, but it has never really been the primary motivation for the development of the projects[3].

In this article we will investigate if learning of science content (or scientific knowledge) can be observed during participation in a selection of Zooniverse projects by testing participants on their science knowledge. We consider five science based Zooniverse projects: Galaxy Zoo, Planet Hunters, Snapshot Serengeti, Seafloor Explorer and Penguin Watch. A summary of these projects is given in Table 1.

**Table 1: Summary of Citizen Science Projects Analysed here.**

| Project | Launch date | Topic | Summary of Task |
|---|---|---|---|
| Galaxy Zoo (www.galaxyzoo.org) | GZ4 launched 11th September 2011. (GZ1 launched 7th July 2007) | Astronomy | Decision tree classification of features seen in images of galaxies from a variety of large astronomical surveys (Lintott et al. 2008). We use data here only from GZ4. |
| Planet Hunters (www.planethunters.org) | 16th December 2010 | Astronomy | Marking of the dips possibly caused by extra solar planets passing in front of stars on graphs of star brightness versus time obtained by the NASA Kepler Satellite (Fischer et al. 2012). We use data here only from the first phase. |
| Snapshot Serengeti (www.snapshotserengeti.org) | 11th December 2011 | Ecology (nature) | Identification of animals in images taken when they set off camera traps run by the University of Minnesota Lion Project in the Serengeti National Park, Tanzania. |

[2] See www.zooniverse.org/publications for a full listing of all publications resulting from Zooniverse projects.

[3] The philopsophy of Zooniverse projects is described at: www.citizensciencealliance.org/philosophy.html

| | | | |
|---|---|---|---|
| Seafloor Explorer (www.seafloorexplorer.org) | 13th September 2012 | Ecology (nature) | Identification of sea animals in images taken with HabCam (Habitat Mapping Camera System), a cable imaging system which dives below a ship to take 6 images a second of the seafloor. |
| Penguin Watch (www.penguinwatch.org) | 17th September 2014 | Ecology (nature) | Counting and classifying penguins in images from cameras overlooking colonies of Gentoo, Chinstrap, Adélie, and King penguins in the Southern Ocean and along the Antarctic Peninsula (run by the Penguin Lifelines project). |

Participants in all Zooniverse projects are engaged in helping with academic research, but questions remain over how much they learn about the scientific method, or the topic of their chosen project(s) during the process. All Zooniverse project websites have sections labelled either "Science" or "About" which provide some basic explanation of the scientific goals behind the project, and there is a collection of educational materials for many Zooniverse projects hosted at www.zooteach.org. However the majority of projects include very little in the way of formal educational material, and while the Zooniverse management encourages science teams to engage with the volunteers (e.g. via the custom "Talk" software, social media or blogs), and volunteers may also learn from each other via Talk, there is significant variation in the levels of actual engagement activity. The projects in Table 1 represent a range of levels of public engagement activity.

There is evidence to suggest that Zooniverse projects can be successful scientifically without significant public engagement, but that they are unlikely to be a success at public engagement without scientific output (Cox et al. 2015). In that work, public engagement success is measured via a combination of the science team activity (in social media and blog posts, and through engaging citizen scientists in the publication process) and volunteer activity (number of volunteers, length of engagement). Among the projects considered here Cox et al. (2015) report a range of success in both public engagement (Galaxy Zoo, Planet Hunters, Snapshot Serengeti and Seafloor Explorer ranked 4th[4], 5th, 2nd and 8th respectively, out of a total of seventeen projects) and scientific impact (ranked 9th, 2nd, 7th and 13th). Our fifth project, Penguin Watch had not launched at the time of that analysis.

The opportunity to learn about science via hands on experience was one of the motivations explored by Raddick et al. (2010) to explain participation in Galaxy Zoo. They found that it contributed to the most important motivation of a small, but non-zero fraction of Galaxy Zoo participants (10%), and was the fourth most frequently mentioned of the "most important motivations", (after interest in astronomy, a desire to contribute to science, and amazement over the vast scale of the Universe). An additional 2% of users indicated that the main reason they use the site to teach others about the science of astronomy.

[4] In Cox et al. (2015), Galaxy Zoo is split into four sub-projects; Galaxy Zoo 4, which we study here, ranked 4th; Galaxy Zoos 1, 2 and 3 are ranked 3rd, 1st and 7th respectively.

It is well known that learning can be intrinsically rewarding (i.e. that some people are motivated to learn for the sake of learning). Making learning fun is also known to make it more effective (e.g. Malone and Lepper, 1987). Aspects of gamification in Zooniverse projects, and the role of fun in the motivation of volunteers are discussed in Greenhill et al. (2014). Evidence was found that gamised activity motivates volunteers to participate. This suggests that aspects of learning linked to the fun had by participants in citizen science projects may be worth exploring further, but that is beyond the scope of this article.

Previous work has found a correlation between astronomical content knowledge and length of participation in two Zooniverse Astronomy Projects (Galaxy Zoo and Moon Zoo; Prather et al. 2013). Another study used measures of the change in language of Zooniverse users on "Talk" between the first and last 10% of posts to demonstrate learning (Luczak-Roesch et al. 2015). That study considers four of the projects discussed here (Galaxy Zoo 4, Planet Hunters, Seafloor Explorer and Snapshot Serengeti), finding that the volunteers in the two astronomy projects showed a much smaller vocabulary shift than those in Seafloor Explorer and Snapshot Serengeti. This might either indicate that Zooniverse users were already familiar with astronomy at the start of the study period (e.g. the tracking began with the launch of new Zooniverse "Talk" software 2012, after 5 years of operation of Galaxy Zoo), or that there is a larger influx of new users into the astronomy projects compared to the ecology projects.

**Objective**

We ask in this article if there is evidence that participation in online citizen science projects can stimulate scientific knowledge learning even in the absence of direct educational motivation for the project design. We will test the hypothesis that while participating in online citizen science, volunteers develop their knowledge about both the science specific to the project they are involved in, as well as becoming more knowledgeable about a set of science topics unrelated to their project (which we shorthand as "general science" hereafter). Finally we will consider the role that public engagement between the science team and volunteers has on the science content learning behaviour of the volunteers. Modern thinking about science learning asks us to remember that a scientifically literate person not only retains a sets of scientific facts, but also understands the processes and context of science (e.g. Sturgis & Alum 2004, Lang et al. 2006, Wynne 2006, Bauer et al. 2007). In this study, we explicitly measure only the learning of scientific content (i.e. knowledge) - just one aspects of full scientific literacy. However the acquisition of scientific knowledge is one of the characteristics of a scientifically literate population (Miller 2001), which makes it a valid (albeit partial) measure of science learning. A study which focuses on the development of the online citizen scientist's understanding of the scientific process, and the contexts and institutions in which science occurs is beyond the scope of this work.

**Methods**

1. **Survey**

As part of the VOLCROWE (Volunteering and Crowdsourcing Economics) project, we have conducted a survey of users in the five Zooniverse projects described in Table 1. This survey was initiated with the goal of studying the motivations of Zooniverse participants (e.g. Cox et al. 2015), but also included a basic general science knowledge quiz and a project specific science knowledge quiz[5].

A pilot version of the survey was run in September 2014. This was designed to measure the response rate among Zooniverse users with different activity levels, and across the different projects, with the goal of constructing a final sample that was representative of the engagement patterns of all volunteers. It was expected (and observed) that the response rate from poorly engaged users would be much lower than those more engaged. In the pilot survey we measured a response rate that was seven times higher amongst the most engaged users (10.3%) compared to the least engaged (1.4%). Answers to the pilot survey are not used in this analysis.

The final survey ran from March 30, 2015 to April 6, 2015, and was sent to 163,686 volunteers. We collected 2737 responses (an average response rate of 1.7%, not atypical for this kind of online survey, e.g. Anderson & Aydin 2005). After removing some incomplete responses, the final sample available for analysis contains 1921 volunteers. The breakdown of this total between the five projects discussed here is: 574 responses from volunteers in Galaxy Zoo; 475 from Planet Hunters; 398 from Penguin Watch; 309 from Seafloor Explorer and 165 from Snapshot Serengeti. Making use of the pilot survey data on expected response rates, we invited a much larger number of the least engaged volunteers compared to the more engaged in order to obtain a representative sample (but this also has the effect of lowering the average response rate). No previous survey undertaken with Zooniverse users has taken such steps to ensure the representative nature of their sample across the range of volunteer engagement. As discussed below in the Results Section (e.g. Table 2 and Figure 3) this effort was largely successful, with only the extreme end of the least engaged volunteers (those who contributed just no more than 2 classifications) being slightly under-represented (they make up 13% of all volunteers, but 9% of our survey respondents).

Zooniverse users may contribute to multiple projects, and the cross over between projects can be significant (Luczak-Rosch et al. 2014). In what follows the engagement (e.g. classification count, length of participation) for the project covered by the respondent's survey answers only is included, ignoring their possible contributions to other projects. No individual (as identified by a unique Zooniverse username) was invited to participate in more than one survey. So for example a volunteer who classifies on both Galaxy Zoo and Penguin Watch, but was invited to answer the survey for Penguin Watch (i.e. with the science quiz tailored to penguin related questions), would only have their Penguin Watch classifications counted as a measure of their engagement with citizen science.

Participants responded to both a general science knowledge quiz and a project specific science knowledge quiz. These science quizzes were developed in consultation with a panel of members of the science teams[6] from across the Zooniverse. Each set of questions consisted of a series of science-related images and participants were asked to state in a free-form text box what was shown in each image. The set of images in the science quiz were specifically built to assess knowledge of facts relating to both general science and specific projects. Each set was designed to contain a mixture of easy and hard questions (Table 3 in the Result Section demonstrates the extent to which this was successful). The project specific questions were designed to test a range of very commonly encountered objects in each project as well as objects more rarely encountered. As an example we include the images used for the general Science Quiz and the Galaxy Zoo and Snapshot Serengeti Project Specific Quizzes in Figure 1. Each set (i.e. the five different project specific sets, as well as the "general science" set) contained five questions and answers were marked on a three-point scale ranging from a basic response (e.g. identifying an image in the middle row of Figure 1 as a galaxy) to advanced answers (e.g. identifying the animal at the lower right as a Reedbuck, or using advanced scientific language in the answer). We reproduce the full answer key used in marking in Appendix A. This key was developed by one of us (KM) in consultation with the panel of Zooniverse scientists. Total scores range from 0 (participant could identify no images correctly) to 15 (participant answered all questions as correctly as possible and using scientific terminology).

[5] The survey can be viewed in full at www.volcrowe.org/survey

[6] All subscribers to the internal "Zooscientists" mailing list were invited to comment on the quiz and answer key.

The VOLCROWE survey was designed primarily as a test of models of user engagement and motivations. These custom designed project specific and general science content knowledge quizzes were included as a potential control variable for those works. As a survey of an online population, it was decided that a novel image based (and therefore difficult to "Google") set of questions needed to be developed. The downside of this technique however, is that (unlike Brossard et al. 2005 who explicitly chose a nationally calibrated general science knowledge instrument in their study of citizen scientists) we will not be able to place the general scientific content knowledge of our survey sample in context with the wider scientific content knowledge of the population.

In order to be able to assess the significance of any conclusions we draw from volunteer scores on the visual science quizzes we must first consider the validity of the quizzes, and assess how well they measure what we intended them to measure. In this work we intend to use the visual science quizzes to measure the scientific content knowledge (either general, or specific to the relevant Zooniverse project) of volunteers in a Zooniverse project. We want to be able to interpret quiz scores such that higher scores imply a volunteer who is more knowledgeable about science, and we want to test if on average volunteers who have spent longer on their project have higher scores.

We assessed the "Face Validity" of the quiz following the method described in Barder et al. (2007). Project specific quizzes were assessed by inviting professional scientists behind each project to comment on the face validity of the set of images. We also looked at image types commonly (and less commonly) discussed on the project Talk interface. The content validation of the quiz was also assessed via consultation with the panel of Zooniverse scientists. We discuss in the Results Section below the range of answers to the quizes, their difficulty, discrimination and discuss their internal reliability.

**Figure 1: Example Science Quiz Images. Top Row: General science quiz. Middle: Galaxy Zoo science quiz. Lower: Snapshot Serengeti quiz. The instructions asked participants to identify the object in the image. Answers are listed in Appendix A.**

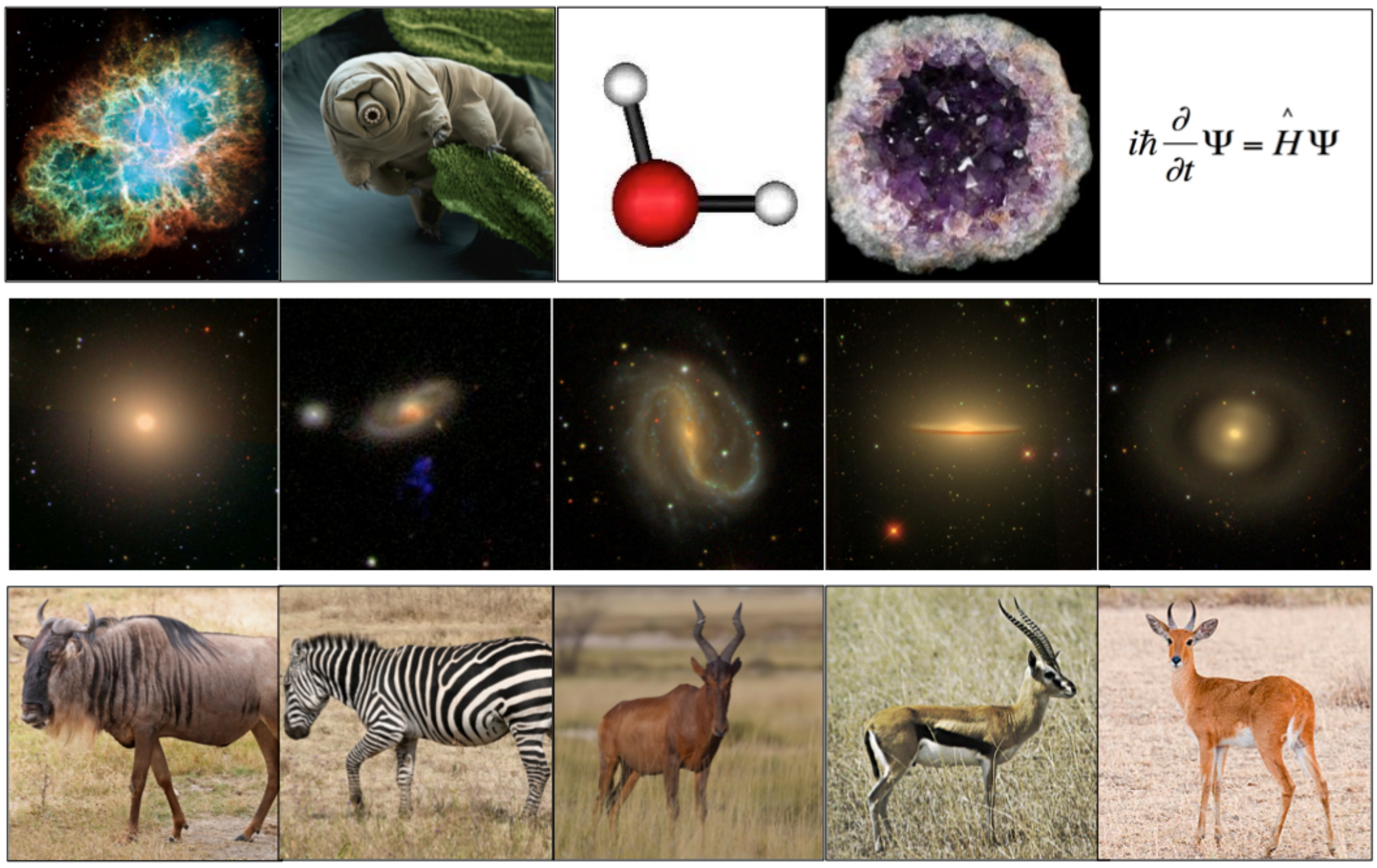

To test how science knowledge correlates with measures of participation in online citizen science we need to be able to control for the influence of circumstances that are known to influence science knowledge, such as the level of education, age, gender or other factors. The control variables we use here are selected from the VOLCROWE survey in accordance with numerous empirical studies across a range of, mostly, but not exclusively Anglo-American populations, which have revealed factors which correlate with science knowledge (Day and Devlin (1998); Hayes & Tariq (2000); Bak (2001); Sherkat (2011); Hayes (2001); Roten (2004); Sturgis & Allum (2004); Gauchat (2011)). These control variables are gender, age, ethnicity, community type (specifically rural or urban), educational level (as measured by ISCED categories) and if the highest qualification is in science and extent to which respondent agrees that religion is important in their life (on a Likert scale). The degree to which religion is important to a person has been found to correlate with scientific engagement in Europe (e.g. Sturgis & Alum 2004 found in a survey of the British population that being non-religious correlated positively with scientific knowledge and attitudes towards science) as well as in the USA. Our literature review includes surveys of populations in Britain, the US, Switzerland, Canada, New Zealand, Norway, The Netherlands, Germany, and Japan). We are confident this is reasonably representative of the population of Zooniverse volunteers.

Finally, we construct two factor scores we call "attitude to science" and "opinion on science learning". These were based on questions appearing in the Volunteer Functions Inventory (Clary et al. 1996), tweaked to be more contextually relevant. Both factors are constructed following principle component analysis of a set of three responses (as described below). The first factor is aimed at measuring the respondent's attitude to science based on responses to the following three questions:

1. the extent to which respondent agrees that participating in Zooniverse allows them to contribute to a cause that is important,
2. the extent to which respondent agrees that the sciences receive adequate funding through taxation,
3. the extent to which respondent agrees that all of society benefits from scientific research.

The second factor is designed to reflect the respondent's opinion on if they are learning while participating in Zooniverse projects though answers to if their participation in Zooniverse

1. lets them learn through direct, hands on experience of scientific research,
2. allows them to gain a new perspective on scientific research,
3. helps them to learn about science

(the results of these questions on learning are discussed further in Results Section 3).

## 2. Econometric Methodology

The goal of our analysis is to model the effect of participation in online citizen science projects upon individual science knowledge. In order to control for the range of other factors that may relate to science content learning outside of engagement with citizen science, we use a technique which allows us estimate a number of multiple regressions - that of standard Ordinary Least Squares (OLS) regression.

The technique is widely used in econometrics in order to test hypothesised causal relationships between two or more variables. In this case, we estimate the relationship between science knowledge (which we treat as the predicted variable) and participation in Zooniverse projects (predictor variable), while explicitly controlling for and holding constant other observable factors that might also relate to scientific content knowledge. The model aims to find a linear combination of the range of independent variables that best predict the science knowledge ($K_i$) of individual i.

The basic regression takes the form:

$$K_i = \beta_0 + \beta_1 P_i + \sum_{j=2,n} \beta_j C_{ij} + \mu_i, \qquad (1)$$

where the dependent variable, $K_i$ measures person i's scientific content knowledge (namely the scores from either general or project specific quizzes, entered as a natural log), $P_i$ is a measure of their participation in the online citizen science project (either a length or time, or a number of classifications, and in both cases entered as a natural log), $C_{ij}$ is a vector representing the set of control variables which we consider might affect science knowledge (as discussed above, and see Table B1). The constants, $\mu_i$ parameterize any constant offset. OLS provides a way to fit the β constants by minimizing the sum of the squared residuals between the predicted and measured values of $K_i$.

We fit for a number of control variables in Equation 1, however there remains a possibility that science knowledge may be an endogenous variable (i.e. there might exist an unknown variable which might drive any trends we see between science knowledge and Zooniverse participation), in which case OLS would not be appropriate. In econometrics, endogeneity can be tested for using the Durbin-Wu-Hausman Test[7] to see if the coefficients between OLS and a Two Step Least Squares regression are different. Applying this test, we find no evidence of engdogeneity, therefore conclude the use of OLS is reasonable.

## Results

### *1. Demographics of Survey Participants*

We show the basic demographics of our survey participants in Figure 2. We note that we have grouped answers into the classifications used for the control variables - the survey included a much wider range of possible answers.

### *2. Participation in Zooniverse Projects*

There is a large variation in the amount of engagement volunteers have with Zooniverse projects (Simpson 2014). A large fraction of contributions are provided by a small group of citizen scientists, while most volunteers engage very minimally with the projects. Our sample was constructed to be representative of Zooniverse volunteers across these engagement patterns (see Methods). We summarize here the measures of participation we will use for our survey respondents.

The distribution of classification activity among respondents to the full survey broadly matches the distribution among all volunteers in our five projects (see Figure 3). We provide some statistics of the number of classifications in Table 2.

**Table 2: Statistics of number of classifications provided by survey sample and all active volunteers in the five projects (between signup to project and September 2014).**

| | Mean | Median | Maximum | N<=2 | N>1000 | N>10,000 |
|---|---|---|---|---|---|---|
| Survey sample (N=1921) | 250 | 18 | 58,432 | 174 (9%) | 69 (4%) | 8 (0.4%) |
| All volunteers (N=315,983 | 153 | 18 | 499,006 | 41,398 (13%) | 7594 (2%) | 429 (0.1%) |

[7] The test was first introduced by Durbin (1954) and developed by Wu (1973) and Hausman (1978) thereafter.

**Figure 2: Pie charts illustrating the demographic makeup of the survey**

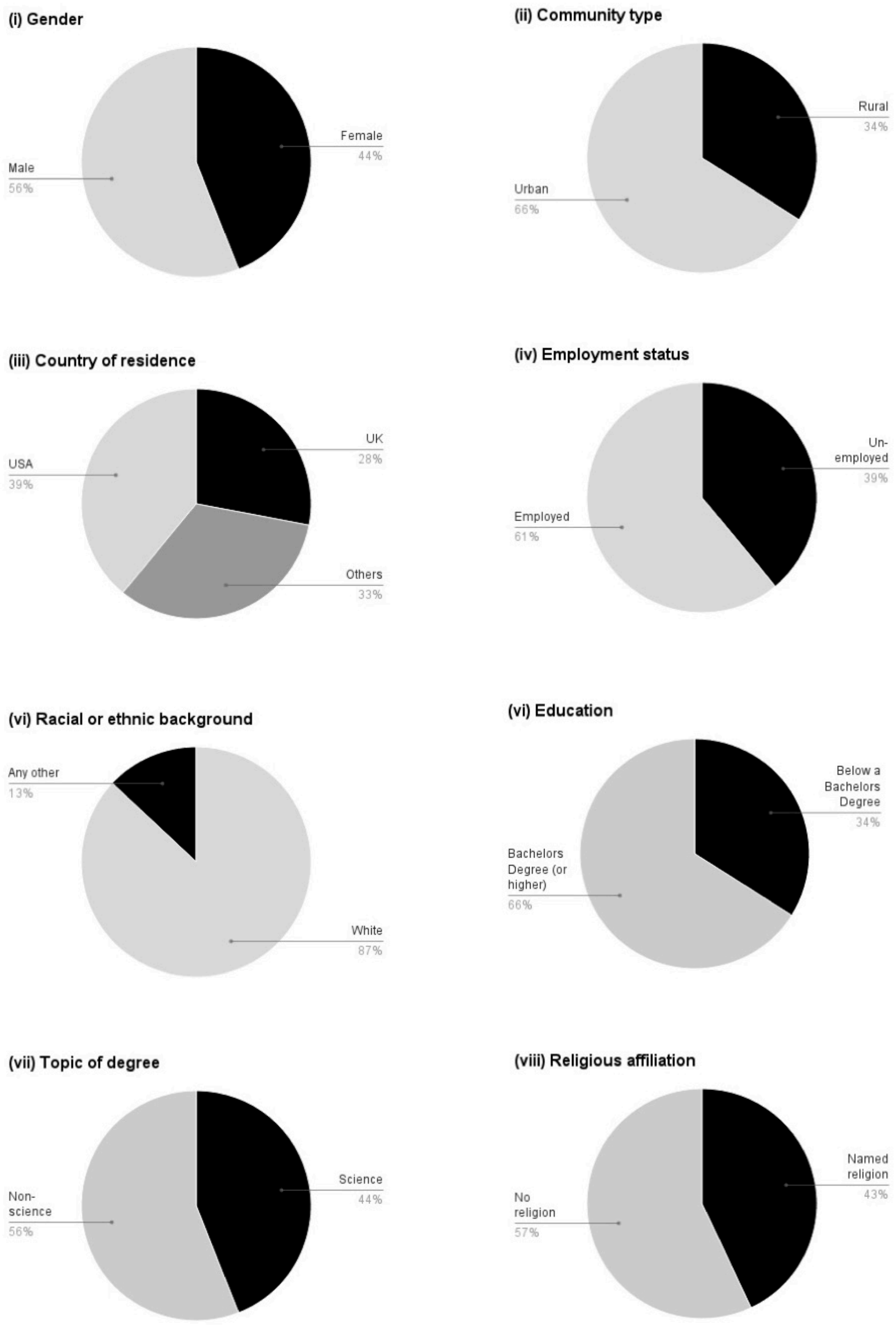

We observe a slight tendency for our survey sample to under-represent volunteers who have contributed two or less classifications, with 13% of all volunteers, and just 9% of our survey respondents in this category. The survey was sent to all potential volunteers in this category, so this is an unavoidable result of low response rates among the most unengaged volunteers.

**Figure 3: Histogram of the number of classifications contributed by the VOLCROWE survey respondents (solid line) compared to the distribution for all volunteers (blue dashed line).**

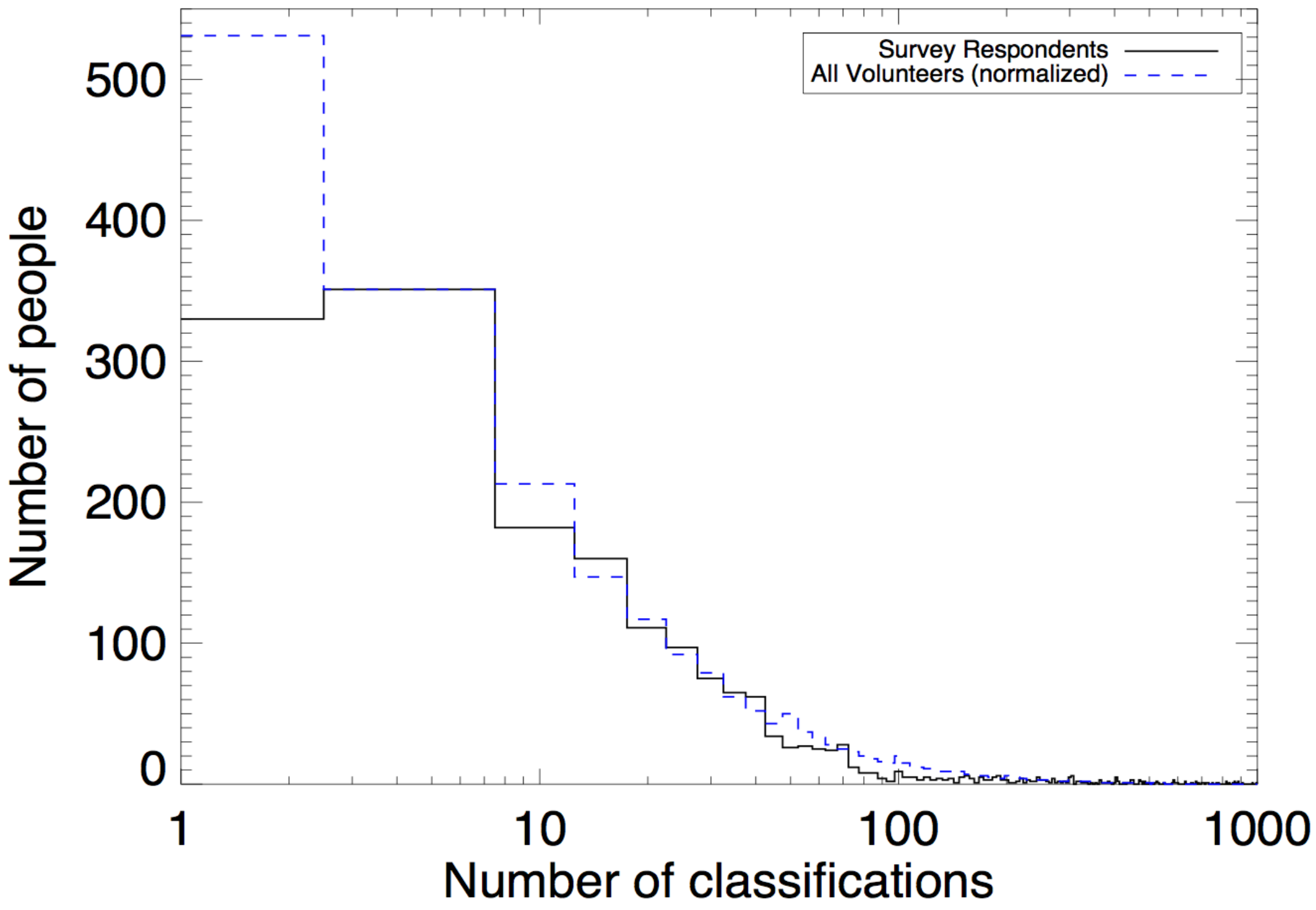

In addition to data on classifications, we also have access to the length of time respondents have been active in the relevant Zooniverse project. We define three durations of interest:

1. *Time since first classification:* Length of time between first classification on the project in question and the survey date (30th March 2015)
2. *Active period:* Length of time between first and last classification to the project in question.
3. *Active days*: The number of unique days in which the respondent supplied at least one classification.

We show the distribution of the length of time since the first classification among our survey sample in Figure 4. This has a flat distribution between very recent sign ups to the Zooniverse (a minimum of 21 days) and a maximum value of 4.3 years, with several noticeable peaks, corresponding to the launch/relaunch of projects considered here (specifically Penguin Watch which launched on 17th Sep 2014, or 197 days; Galaxy Zoo 4 on 11th September 2011 or 933 days; and Seafloor Explorer on 13th Sept 2012, or 931 days) as well as mentions of Zooniverse projects on the BBC Stargazing LIVE program (8th January 2013, or 812 days; 16th January 2012, or 1172 days; and 4th January 2011, or 1547 days).

**Figure 4: Histogram of the length of time since the first classification to 30th March 2015. The peaks correspond to either project launches or mentions on BBC Stargazing LIVE.**

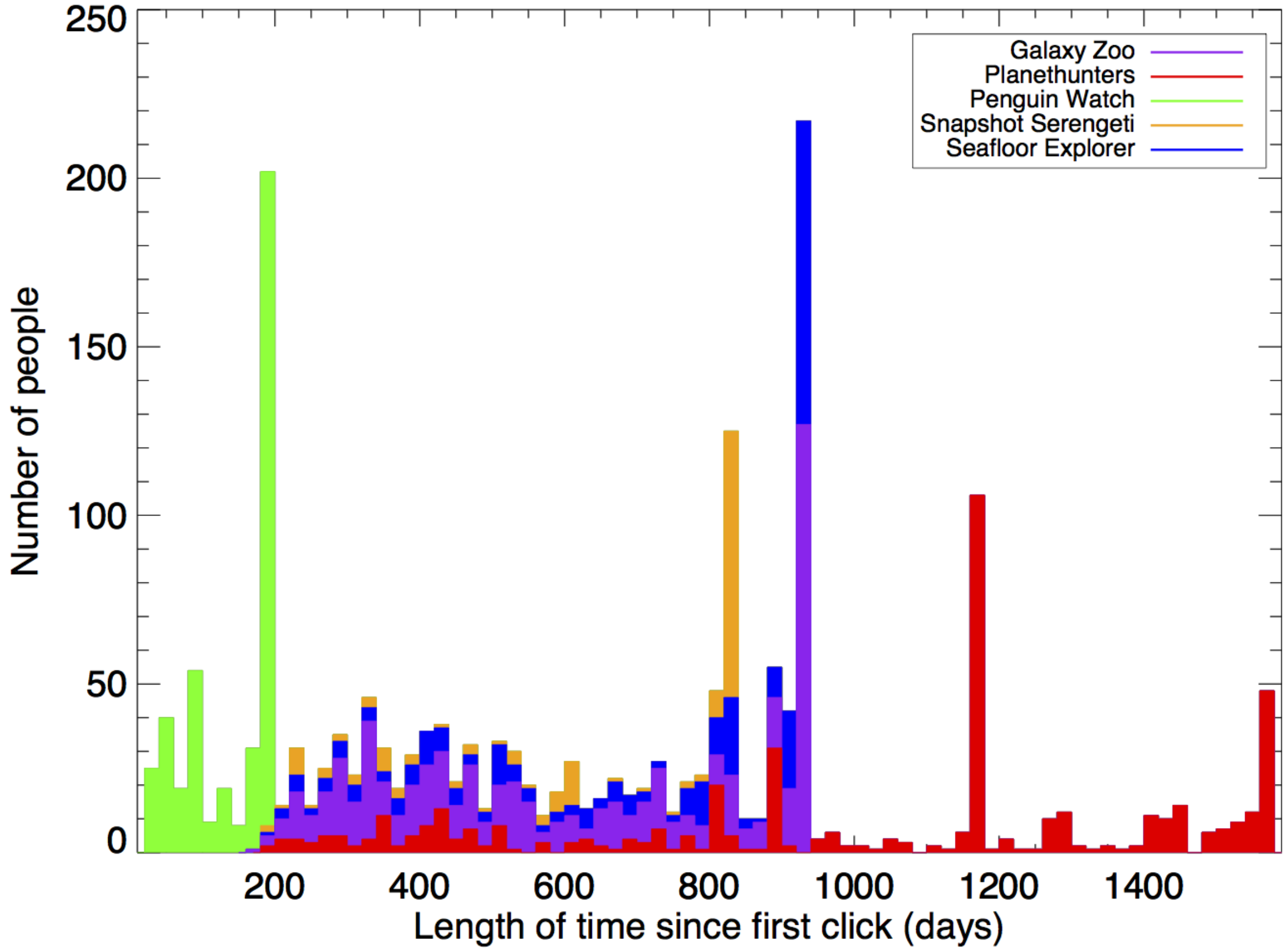

Both the "length of the active period" and the number of "Active Days" have a very skewed distribution, with a large peak of low engagement (ie. a single day), representing people who only ever contributed a small number of classifications in a short time after their first click, and a long tail (to a maximum active period of 3.4 years, and a maximum of 602 separate active days). The mean value of the active period is 2.3 months, and the mean number of active days is 4.6. We provide a full summary of our data on participation in Table B1 in Appendix B (along with descriptive data on all other variables).

*3. Self Reported Science Learning*

Our survey asked participants to think about their motivation for participating in Zooniverse projects. Three of the suggested motivations were related to science learning. We find that among the 1921 respondents, the majority (more than 80%) agree that they feel that their participation is contributing to their scientific literacy. Figure 5 shows pie charts of the answers to three questions about learning in the Zooniverse. As discussed above, these answers are combined in a factor analysis into a single number which measures "opinion on science learning", and which is included as a control in our econometric regressions below.

**Figure 5: Survey responses of participants views on learning while participating in Zooniverse..**

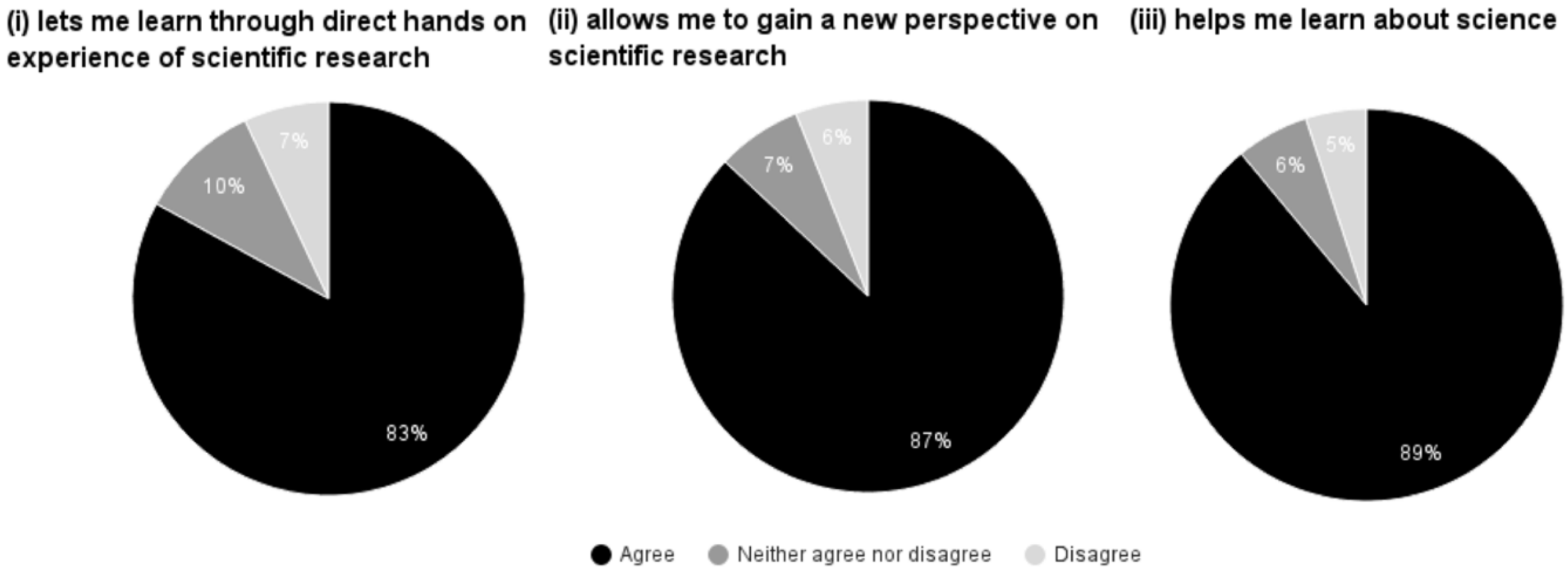

*4. Scientific Knowledge Measured by Visual Quiz*

Here we discuss the results of the visual science quizzes. The average result in both the general science knowledge and project specific science knowledge quizzes was just under 50% (7/15; see Table B1 in Appendix B). All five project knowledge quiz results have similar averages and standard deviations (Table 3), suggesting that the range of questions captured fairly the range of project specific knowledge across the five projects (ie. no one quiz was significantly easier or harder than the others).

Following the method described in Wallace et al. (2011a,b) we also calculate the difficulty (p; the fraction of correct answers) and discrimination (the point biserial, or the correlation between the item score and total score) of each item in each quiz. These values are also provided in Table 3. We note that in order to do this we create in addition to the full score (which ranged from 0 to 3 for each item as described in the Methods Section, and with the answer key given in Appendix A) a binary score (ie. simply correct or incorrect).

**Table 3: The Mean and Standard Deviation, difficulties and discrimination of items on the visual science quizzes.**

| | | General Science | Galaxy Zoo | Planet Hunters | Penguin Watch | Seafloor Explorer | Snapshot Serengeti |
|---|---|---|---|---|---|---|---|
| Mean ± σ | 1 | 1.31±1.05 | 0.94±1.16 | 0.69±1.27 | 1.07±1.43 | 2.70±0.90 | 1.86±0.52 |
| | 2 | 1.05±1.24 | 0.89±1.03 | 2.11±1.37 | 0.84±1.35 | 1.34±1.47 | 2.91±0.52 |
| | 3 | 1.57±1.04 | 1.70±1.07 | 2.67±0.94 | 1.25±1.48 | 0.04±0.33 | 0.56±0.80 |
| | 4 | 1.98±1.13 | 0.91±1.25 | 1.09±1.32 | 1.10±1.45 | 1.92±1.19 | 1.57±1.28 |
| | 5 | 1.13±0.78 | 0.72±1.28 | 1.00±1.24 | 1.01±1.42 | 0.29±0.60 | 0.48±0.91 |
| All | | 7.04±3.15 | 5.17±3.95 | 7.57±3.68 | 5.27±4.07 | 6.29±2.57 | 7.33±2.55 |
| Difficulty | 1 | 0.76 | 0.49 | 0.23 | 0.36 | 0.90 | 0.76 |
| | 2 | 0.45 | 0.44 | 0.70 | 0.28 | 0.46 | 0.97 |
| | 3 | 0.78 | 0.74 | 0.89 | 0.42 | 0.02 | 0.39 |
| | 4 | 0.83 | 0.35 | 0.42 | 0.37 | 0.83 | 0.62 |
| | 5 | 0.86 | 0.24 | 0.41 | 0.34 | 0.23 | 0.27 |
| Discrimination (Point biserial) | 1 | 0.46 | 0.74 | 0.60 | 0.66 | 0.38 | 0.62 |
| | 2 | 0.59 | 0.72 | 0.61 | 0.53 | 0.73 | 0.27 |
| | 3 | 0.52 | 0.80 | 0.53 | 0.72 | 0.31 | 0.42 |
| | 4 | 0.43 | 0.71 | 0.66 | 0.63 | 0.50 | 0.62 |
| | 5 | 0.20 | 0.61 | 0.57 | 0.30 | 0.31 | 0.41 |

We explicitly aimed to have a range of difficulties (p) in our quiz items. This has generally worked; the Penguin Watch quiz images have the most similar difficulties, spanning $0.28<p<0.42$. A a general rule p-values should not differ too much from 0.5, with a range of $0.2<p<0.8$ considered most desirable (Bardar et al. 2007, Wallace et al. 2011b). Nearly all of the items in our quiz fall into the range, the only exceptions are items 1 and 3 in the Seafloor Explorer quiz (item 1 - a starfish - is almost always correctly identified, while item 3 - an anemone - almost never), and item 2 in the Snapshot Serengeti quiz (almost everyone can correctly identified a Zebra). The discrimination of items is calculated using the point biserial (the level of correlation between a correct answer in a given item and the overall quiz score). We find that all items fall into the acceptable range, being larger than 0.2 (indicating a positive correlation, Wallace et al. 2011b). It would also be typical to also report the Cronbach's-α values here. This statistics aims to measure the internal consistency of test scores, such that it is close to one when the items are highly covariant, and close to zero when they are not. However each of our quizzes has just N=5 items, which is too low for this statistic to be useful. Overall we are confident that this analysis demonstrates that the quiz items are successful at discriminating between volunteers with a range of scientific knowledge levels.

We find that there is a broad positive correlation between general science knowledge and project specific science knowledge (Figure 6). Respondents were slightly more likely to do well on the general science questions and poorly on the project specific questions than vice versa (e.g. as you can see in the plot there are no volunteers with very high project scores and very low general science scores, but there are a handful of volunteers with very high general science scores and very low project specific knowledge).

**Figure 6: Density plot of Science Quiz Results for General Science and Project Specific Science. A one-one line is over-plotted to guide the eye.**

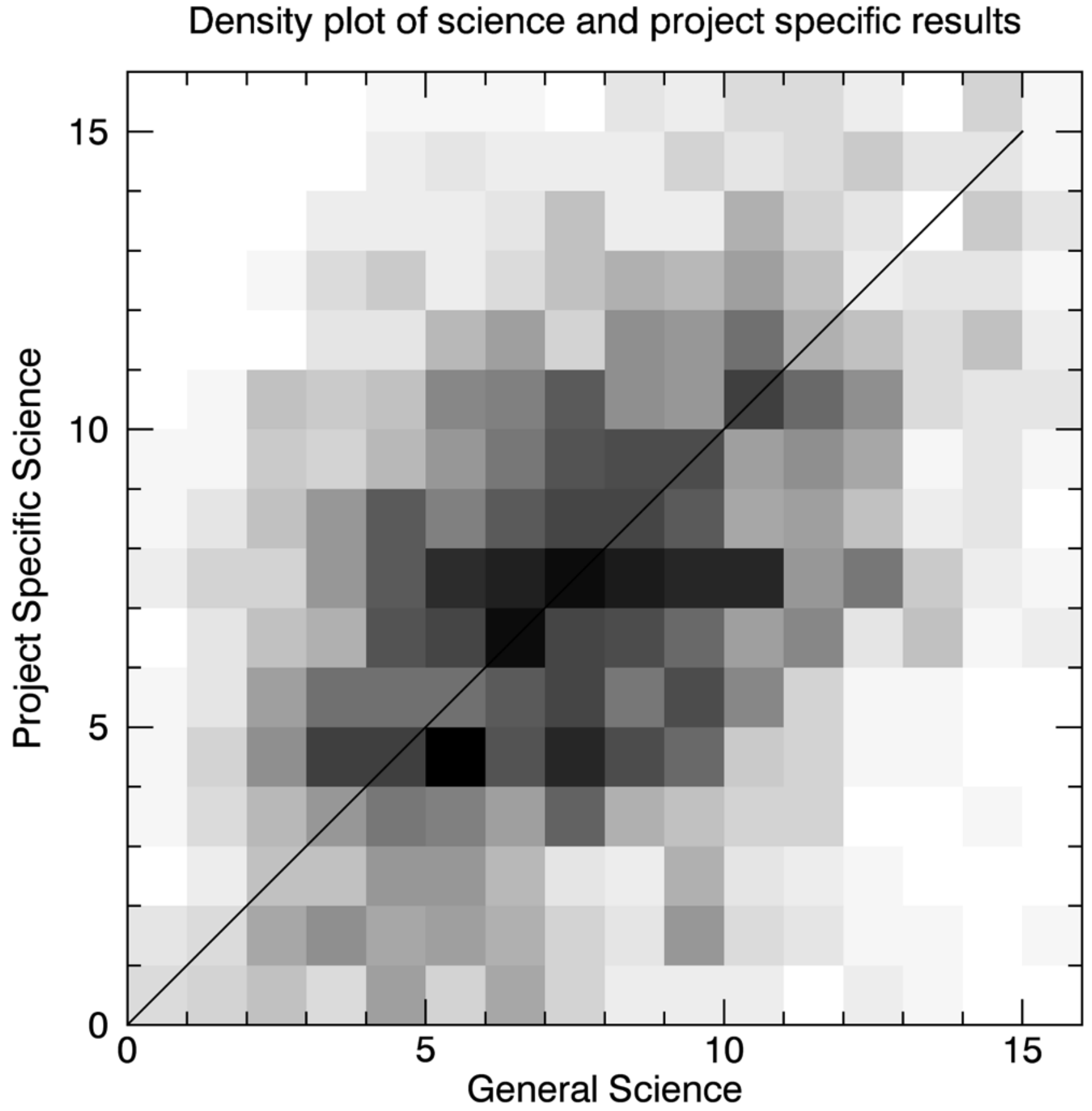

*5. Can Participation Increase Science Knowledge?*

Here we report on results from a set of OLS regressions of the survey data designed to investigate the relationship between science knowledge and participation in citizen science while holding other relevant variables (such as education and science qualifications) constant.

1. *General Science Knowledge*

We start by testing how the general science knowledge score of the survey respondents depends on engagement with Zooniverse projects (controlled for other factors known to correlate with science knowledge as discussed above). We find that no measure of engagement demonstrates a significant relationship with this score. We also find there is no relationship between the factor score measuring 'Opinion on Science Learning' and performance in the general science quiz; i.e. respondents who indicate that they believe they are learning as a result of their participation in Zooniverse projects do not perform any better in the general science quiz than those who do not. Taken together these results imply a null result – we can find no evidence that general science knowledge as measured by our quiz is linked to participation in Zooniverse projects.

We do find a significant heterogeneity in performance across the five projects, with participants in the astronomy-related projects generally observed to perform better compared with participants in the ecology-related projects. Also, as was expected, we find a positive correlation with the factor score measuring 'Attitude to Science' - respondents who regard science more positively tended to also know more about science in general. (See Table B2 in Appendix B for the full regression results).

*2. Project Specific Science*

We now test if participants learn project specific science while engaging in citizen science, by using the scores in the project specific science quiz as the dependent variable. We find that three out of the four measures of engagement show a positive and significant (at 99% level) association with this score (see Table B3 in Appendix B for the regression results). The engagement measure of "time since first classification in the project" is found to not correlate significantly with the score. A doubling of the active period (defined as the length of time between the first and last classification on the project) leads to just a 1% improvement in score, while doubling the number of classifications done leads to a 4% improvement in score, and doubling the number of active days spent participating in the project associates with an 8% increase. Taken together this suggests that it is active engagement (i.e. more frequently visiting the sites and contributing classifications) that is associated with an improvement in science knowledge.

It is not easy to demonstrate the extent to which this relationship is causal or merely reflects correlation. One might equally assume that respondents who have a higher knowledge of the project specific science classify more as assume that this result demonstrates learning is happening. In an attempt to address this concern, we perform a further regression of project knowledge adding our measure of general science knowledge as a control variable. We have demonstrated above the lack of correlation between engagement and general science knowledge, and have reported the broad correlation between general and project specific science. Therefore, we can use performance in the general science quiz as a way to control for 'baseline' scientific knowledge of participants (e.g. a proxy for their science knowledge before joining the project). After introducing this control, the coefficient estimates between project specific knowledge and measures of active engagement remain significant and positive at the 99% level (and very similar levels of score increase as reported above – see Table B4 in Appendix B). We suggest this is evidence that learning is taking place.

We also note that once general science knowledge is controlled for, the correlation between projects specific knowledge and 'Opinion on Science Learning' becomes larger (now significant at the 90% level). This is suggestive that respondents who feel they are learning as part of their engagement with Zooniverse projects are performing better in the project specific science quiz.

All of the above correlations show evidence of project-level heterogeneity in performance, particularly between astronomy projects such as Galaxy Zoo and ecology projects such as Seafloor Explorer. The public engagement success of Zooniverse projects was measured in Cox et al. (2015) by a combination of six factors. Of these, two are measures of project activity on blogs, Twitter and Talk (by both volunteers and the science team), while the others are measures of direct collaboration between professional and citizen scientists, the overall size of the project, and the amount of time/classifications volunteers contribute.

Cox et al. (2015) find that Seafloor Explorer is the worst performing of the five projects in this study in their public engagement metrics, while the astronomy projects tended to perform much better. This suggests that participants in projects with lower success in public engagement may be less likely to learn project specific science. We run the OLS regression separately for project specific knowledge in each of the five projects sub-samples, and are able to measure a significant positive correlation between project specific knowledge and the measures of active engagement in each project separately (Table 4).

**Table 4: Relationship Between Project Knowledge and Engagement for five Zooniverse Projects, and their Public Engagement Success Ranking from Cox et al. (2015)**

| Variable | Classifications | Active Period | Active Days | Public Engagement Success Ranking |
|---|---|---|---|---|
| Galaxy Zoo (N=475) | 0.041** | 0.021** | 0.125** | 4 |
| Planet Hunters (N=398) | 0.027* | 0.013** | 0.050* | 5 |
| Penguin Watch (N=309) | 0.108** | 0.027* | 0.162** | N/A |
| Snapshot Serengeti (N=165) | 0.083** | 0.029** | 0.147** | 2 |
| Seafloor Explorer (N=309) | 0.055** | 0.017** | 0.096** | 8 |

Note: We report β from Equation 1. Statistical significance is denoted *=95% level; **=99% level. See Methods section for list of controls included in the regression.

We compare the strength of the relationship with the ranking of "public engagement" success reported by Cox et al. (2015) in Figure 7. This reveals that the higher ranked projects in public engagement success are found to have a stronger association between active engagement and project specific science knowledge. We suggest that this indicates that the project-level heterogeneity in performance is at least partially explained by the different levels of public engagement success of the projects.

We note that as all five projects are hosted by the Zooniverse, they all were developed with the same philosophy, and include the same general level of educational content. We therefore argue that the biggest relevant differences between Zooniverse projects is the involvement and engagement level of the professional science team, rather than any factors related to site design or educational content.

**Figure 7: A plot of the values of the coefficients in the relationship between active measures of participation and scores in project specific science quiz as a function of the public engagement ranking of different projects from Cox et al. (2015). Horizontal lines show the results for the entire survey sample.**

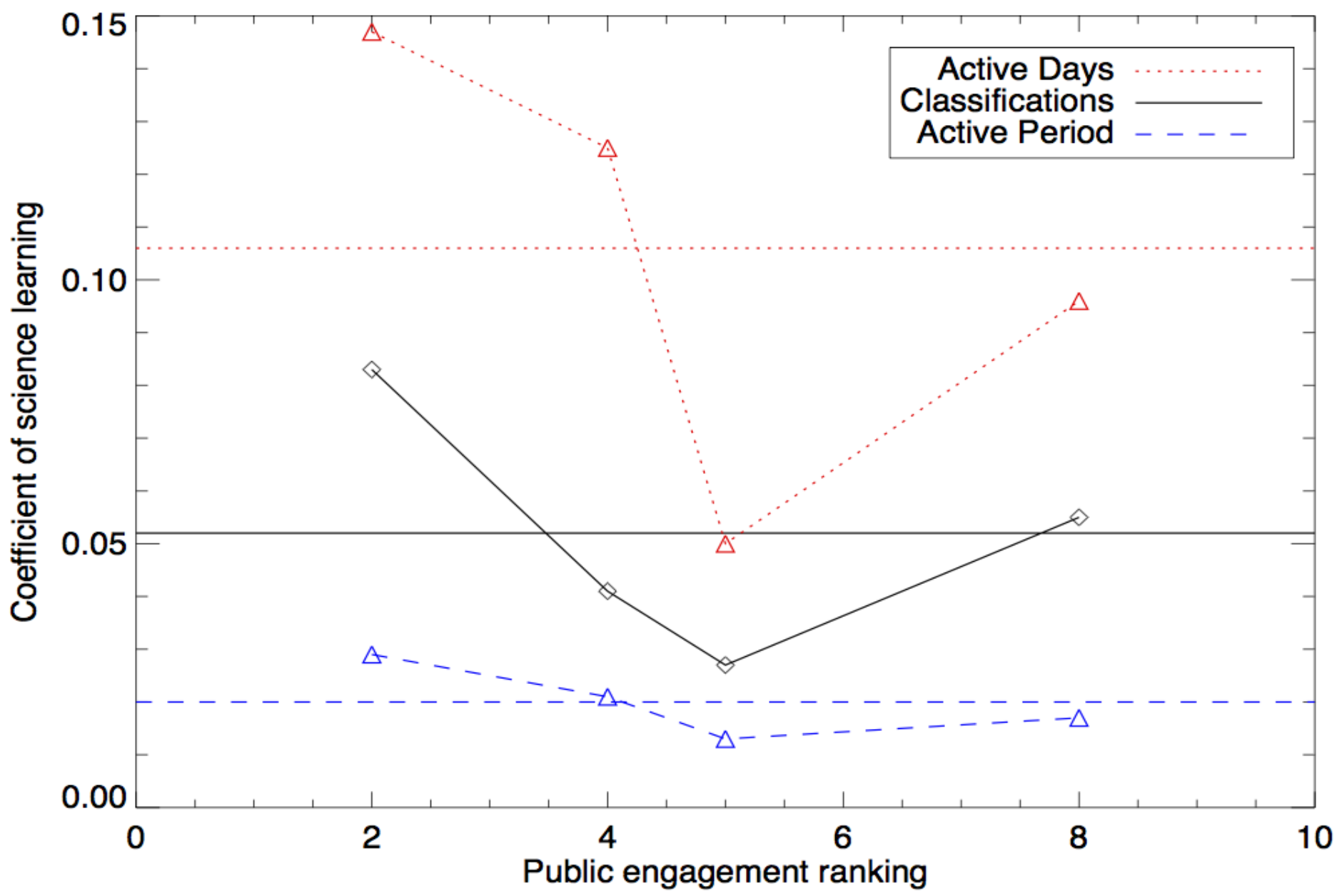

Galaxy Zoo and Snapshot Serengeti, the projects we find have the strongest links between learning and active engagement both perform well in the communication related public engagement factors; while Seafloor Explorer and Planet Hunters, which we find to have weaker links between learning and engagement, perform relatively poorly (Planet Hunters ranks higher overall in Public Engagement by performing well in the direct collaboration metric). This suggests that it is the public engagement in blogs, Twitter and Talk which are driving the link between active participation and learning. This superficially agrees with Wise, Hamman & Thorson (2006) reveal that participants in online communities demonstrate more intent to participate if there is more obvious moderation and DeVries Hassmen et al. (2013), who reconceptualise learning as a part of community engagement in citizen science and find (in a study of Planet Hunters and Seafloor Explorer) that participants are more likely to engage with Talk and other areas of the Zooniverse projects as they move towards more sustained participation. DeVries Hassmen et al.  link this engagement to collaborative learning.

## Conclusions

In this article we have explored how participation in online citizen science is correlated with scientific content (or knowledge) literacy as measured by a set of custom designed image based science quizzes. We have explored users in five different projects hosed on the Zooniverse platform (www.zooniverse.org), namely Galaxy Zoo, Planet Hunters, Seafloor Explorer, Snapshot Serengeti and Penguin Watch. We do this via a survey primarily designed to study the motivations of users as part of the VOLCROWE project (e.g. Cox et al. 2015). We collect answers from 1921 volunteers to a variety of demographic and motivation seeking questions, as well as an image based quiz testing general politics knowledge, general science and project specific science knowledge. While our average response rate is low (1.7%), this was necessary to ensure a large and representative sample of volunteers across the five Zooniverse projects which samples users with different levels of engagement from the least engaged (who have a response rate of 1.4%) to the most engaged (with response rate of 10.3%).

We find that the majority (more than 80%) of users believe that their participation in the Zooniverse projects is contributing to their development of scientific literacy via understanding of the scientific method and science knowledge. In this study we make no test of the former, but we test the development of science content knowledge with our survey.

We find a significant and positive relationship between forms of active engagement with Zooniverse projects (as measured by either length of time since first and last Zooniverse clicks, number of days on which classifications are recorded, or total classifications contributed in a fixed time period) and results in a project specific science knowledge quiz. We find no evidence of such a relationship with general science knowledge. We suggest that this can be interpreted as evidence for science learning while participating in online citizen science, as our results remain consistent even when we control for performance in the general science quiz as a measure of 'baseline' scientific knowledge. This implies, that regular engagement with Zooniverse projects leads to an increase in knowledge of a particular area of science, if not science in general.

It is important to notice a limitation of this empirical test since the OLS model in this paper has a potential endogeniety bias. We have tested for endogeneity, and found it was not significant, even so in future research, it would be worth investigating the endogenous potential of engagement using more advanced econometric methodology, or by removing this concern through running a longitudinal study of citizen science participants rather than a snapshot survey. Another limitation is that our custom designed science knowledge image quizzes mean that we are not able to make a comparison of the scientific knowledge of our survey group with the wider population. Finally, a third limitation of our work is that we test primarily the learning of science content, not scientific literacy more generally (although we ask for self reported answers about more general science learning via participation).

The Zooniverse projects considered here span a range of public engagement behaviours from among the most successful Zooniverse projects in public engagement (Snapshot Serengeti; as measured by Cox et al. 2015 via activity on blogs, Twitter and Talk - by both volunteers and the science team, as well as evidence of collaboration between professional and citizen scientists, the overall size of the project, and the amount of time/classifications volunteers contribute) to a relatively poorly performing project (Seafloor Explorer). We find evidence that participants in the more poorly performing projects learn relatively less through their engagement than those in the more successful public engagement projects, and suggest this is tied to different levels of communication and interaction via blogs, Twitter and Talk in the projects. Any such measurement may however be confounded by the cross over between users of different projects in the Zooniverse (Luczak-Rosch et al. 2014).

Our results imply that even for citizen science project designed primarily to meet the research goals of a science team, volunteers are learning about scientific topics while participating. Combined with previous work (Cox et al. 2015) that suggested it is difficult for projects to be successful at public engagement without being scientifically successful (but not vice versa) this has implications for future design of citizen science projects, even those primarily motivated by public engagement aims. While scientific success will not alone lead to scientific learning among the user community, we argue that these works together demonstrate scientific success is a necessary (if not a sufficient) requirement for successful and sustainable public engagement through citizen science. We conclude that the best way to use citizen science projects to provide an environment that facilitates science learning is to provide an authentic science driven project, rather than to develop projects with solely educational aims.

**Acknowledgements**

The Zooniverse projects in this study were built thanks to support by the Alfred P. Sloan Foundation. VOLCROWE (Volunteer and Crowdsourcing Economics) is a research project funded by the EPSRC in collaboration with the NEMODE network and involving collaboration between the University of Portsmouth, University of Oxford, University of Manchester and University of Leeds. The project is running from September 2013 to September 2016. We acknowledge the helpful comments of Laura Trouille (Adler Planetarium), Edward Prather (University of Arizona), as well as three anonymous referees who helped significantly improve this work.

## Authors

**Dr. Karen Masters** is a Reader in Astronomy and Astrophysics at the Institute of Cosmology and Gravitation, University of Portsmouth. She is the Project Scientist of the Galaxy Zoo Science Team, and the Director of Education and Public Engagement for the Sloan Digital Sky Survey. Email: karen.masters@port.ac.uk

**Dr. Eun Young Oh** is a senior research associate at the University of Portsmouth Business School. She developed expertise in Economics by several years of industrial and academic experiences. Since joining the VOLCROWE team in 2014 Dr.Oh has transferred the skills she honed studying Monetary Policy, DSGE Model, Bayesian Analysis and Panel Data Analyses to the Digital Economy..

**Dr. Joe Cox** is a Principal Lecturer at the University of Portsmouth Business School. His research interests relate to the digital economy and areas such as crowdsourcing, crowd-funding, digital piracy and video games. He is the PI of VOLCROWE.

**Dr. Brooke Simmons** is an astrophysicist researching black holes and galaxies as an Einstein Fellow at UC San Diego (where she recently moved from the University of Oxford). Dr. Simmons is a member of the Zooniverse team, involved in planning and building several different projects within the Zooniverse, as well as interacting with the large community of volunteers..

**Prof. Chris Lintott** is a professor of astrophysics at the University of Oxford, where he is also a research fellow of New College. He is the principal investigator for Zooniverse. His research focuses on the development of sophisticated platforms for citizen science pattern recognition and anomaly detection (as well as galaxy evolution).

**Prof. Gary Graham** is an associate professor of operations and supply chain management at Leeds University, and a visiting research scholar at MIT Future Freight Lab. He has authored three books, 34 research papers, and his grant income (EU, EPSRC and ESRC) totals £1.13m.

**Dr Anita Greenhill**'s research interests are in the areas of Networked Usage of Technology within Community, Organisational and Business settings. Dr Greenhill's research focus on building a better understanding of crowdsourcing and Internet participation; Community usage/resilience and its use of social media including twitter; technologically enable work; spatiality; and Web Usage in Organisations.

**Dr. Kate Holmes** formerly worked as a Researcher in the VOLCROWE project at Manchester Business School and is now working within Research Data Management at The University of Manchester. Her research interests include independent creative enterprise and DIY culture, information and communication technologies, crowdsourcing, and online communities.

**Appendix A: Key to Science Quiz Images Identification**

| Image/Question[8] | Example wrong answers (0 points) | Basic answers (1 point) | Intermediate answer (2 points) | Advanced answer (3 points) |
|---|---|---|---|---|
| 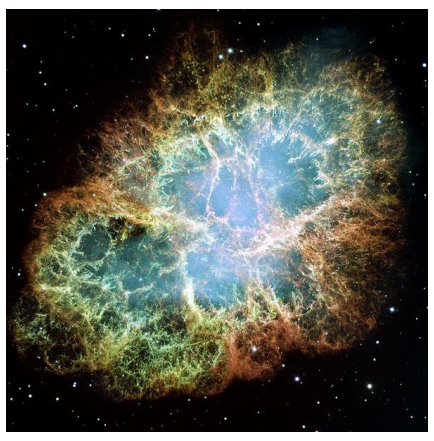 | galaxy,<br>the big bang,<br>constellation | nebula,<br>gas cloud | supernova remnant,<br>exploding star | Crab nebula,<br>M1,<br>Messier 1 |
| 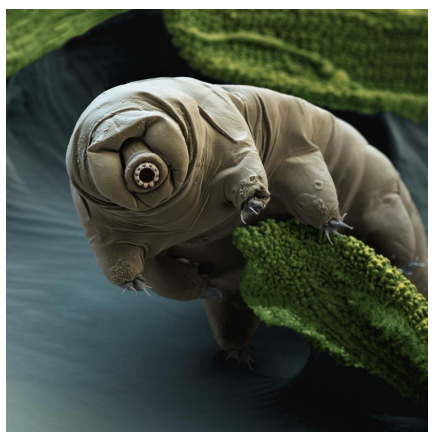 | caterpillar,<br>manatee,<br>eyelash | bug | Tiny bug,<br>Microscopic bug,<br>microscopic animal,<br>Waterbear,<br>microorganism,<br>moss piglet | tardigrade,<br>extremophile, |
| 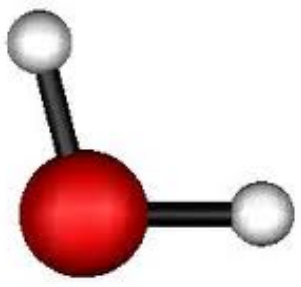 | atom (singular),<br>hydrogen | molecule | water,<br>water molecule | $H_20$ |
| 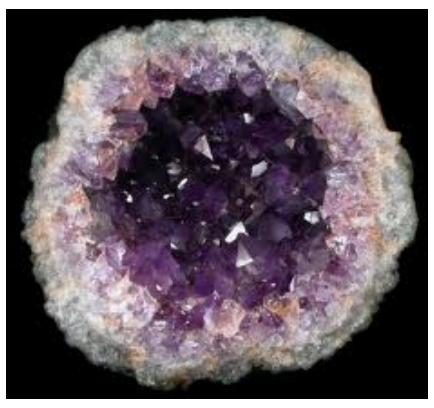 | rock,<br>agate,<br>supernova | crystal | amethyst,<br>quartz | geode |
| $i\hbar \frac{\partial}{\partial t}\Psi = \hat{H}\Psi$ | greek,<br>Hebrew,<br>Egyptian hieroglyphic | formula,<br>math(s),<br>equation | quantum,<br>differential equation | Schrodinger's equation |
| 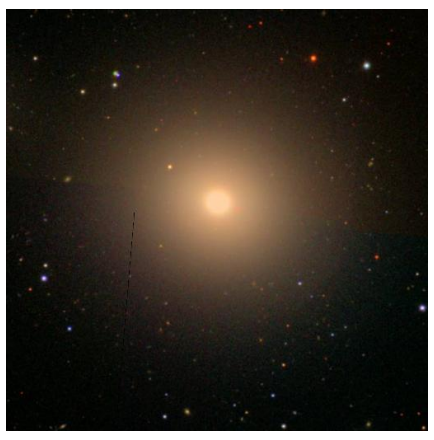 | star | galaxy | smooth,<br>"no spiral arms" | elliptical,<br>NGC 3379 |

[8] If no question is given the default is "Please identify the object in this image."

| | | | | |
|---|---|---|---|---|
| 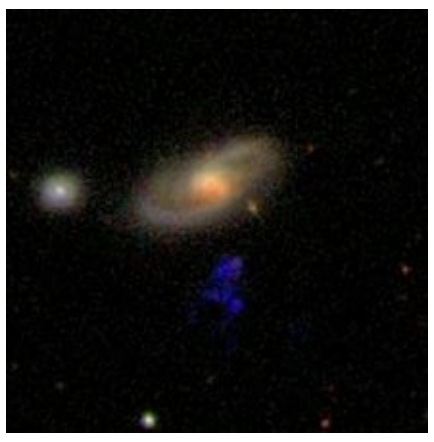 | planet | galaxy | spiral,<br>voorwerp | Hanny's<br>Voorwerp,<br>IC 2497 |
| 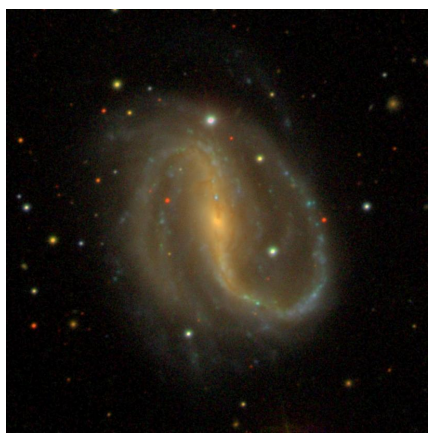 | elliptical | galaxy | spiral | barred spiral,<br>NGC 7479 |
| 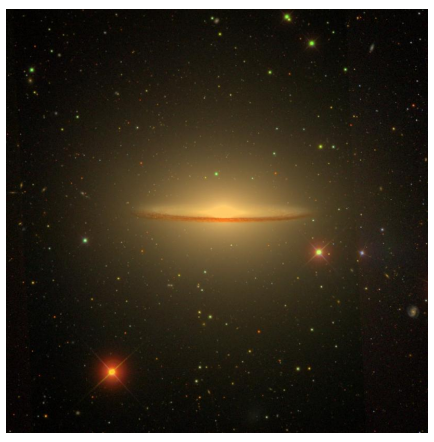 | elliptical | galaxy | disc/disk,<br>edge-on | Sombrero,<br>lenticular,<br>S0 |
| 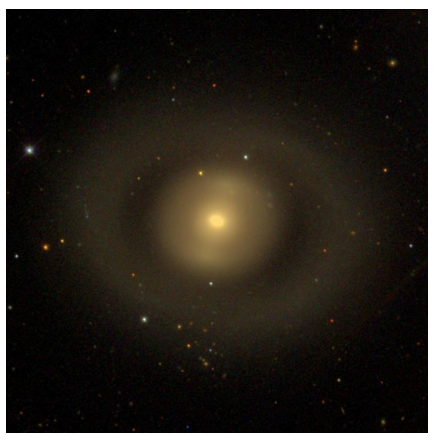 | nova,<br>black hole | galaxy | ring galaxy,<br>barred galaxy | NGC 2859 |
| 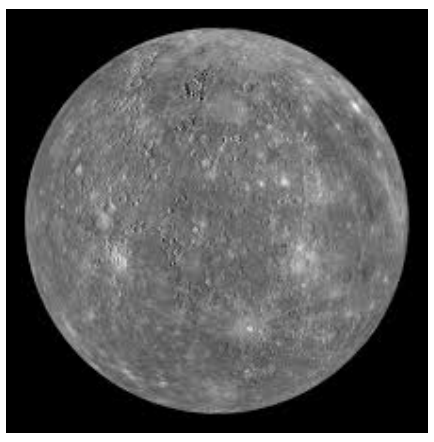 | Moon, Io | planet | rocky planet | Mercury |
| 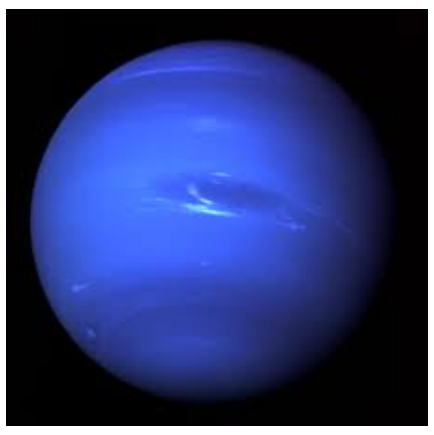 | Uranus | planet | gas giant,<br>gaseous planet,<br>ice giant | Neptune |
| 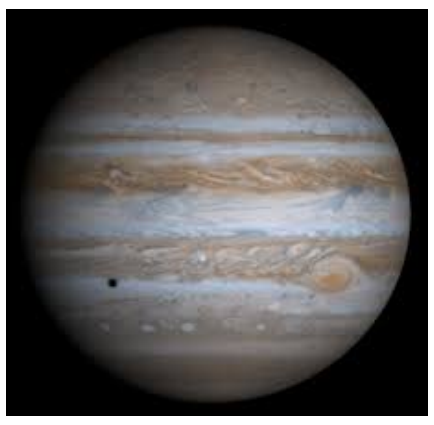 | Mars | planet | gas giant | Jupiter |
| 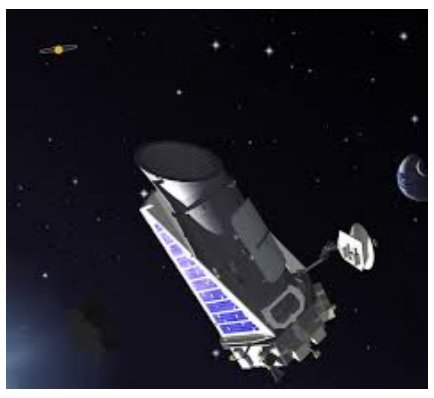 | Hubble | satellite | Space telescope,<br>telescope | Kepler |
| 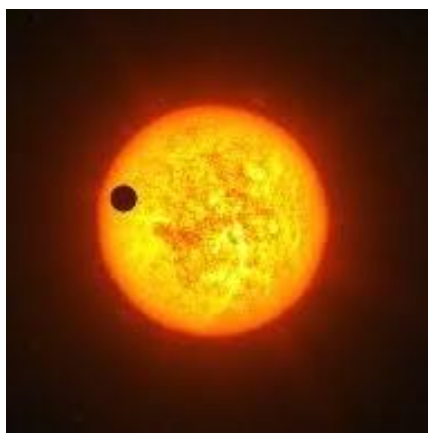 | Sunspot,<br>Moon,<br>Mars transit | Sun,<br>star | transit,<br>eclipse,<br>planet | planetary transit,<br>mercury transit,<br>venus transit,<br>CoRoT-9b ,<br>exoplanet |
| 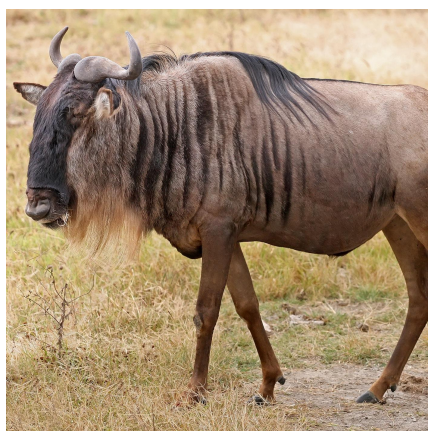 | yak,<br>bison,<br>water buffalo,<br>horse | antelope | N/A | wildebeest,<br>gnu |
| 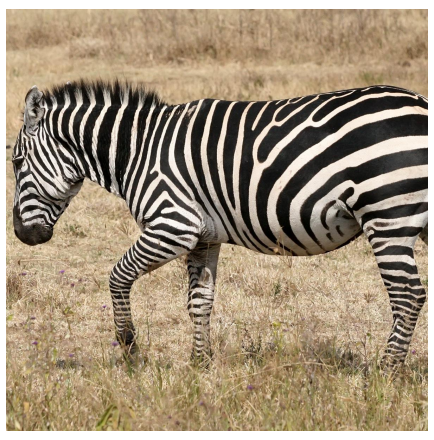 | | zebra | N/A | equid,<br>Crawshay's,<br>Grevy's[9] |
| 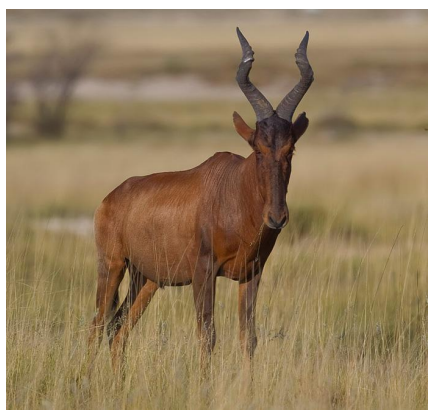 | ibex,<br>oryx,<br>elk | antelope | N/A | hartebeest |
| 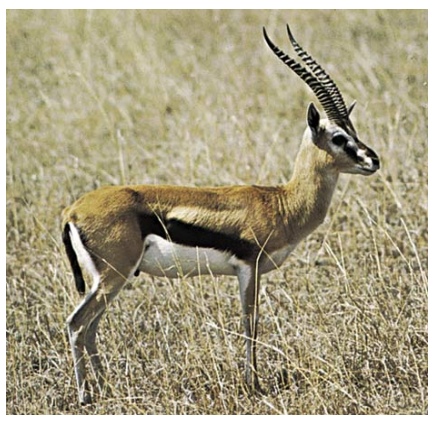 | antelope,<br>impala | N/A | gazelle | Thompson's<br>gazelle,<br>Tommie |
| 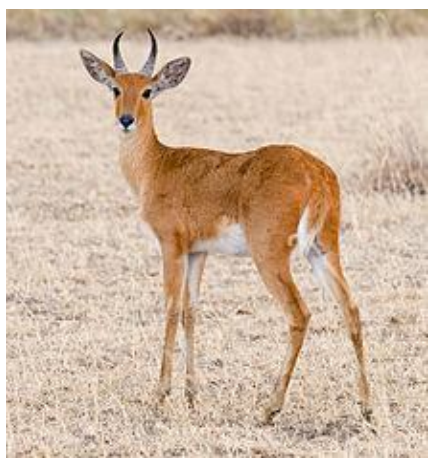 | gazelle,<br>deer,<br>impala,<br>dik dik,<br>springbok | antelope | N/A | reedbuck,<br>redunca,<br>bohor |
| 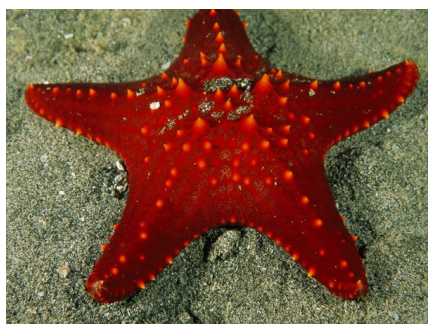 | Stella marina | starfish,<br>sea star,<br>sea-star | N/A | asteroidea,<br>asteroid<br>enchinoderm,<br>bahamian |

[9] We recognise it is impossible for this Zebra to be both Crawshay's and Grevy's, but we give 3 points for the knowledge of the terms.

| | | | | |
|---|---|---|---|---|
| 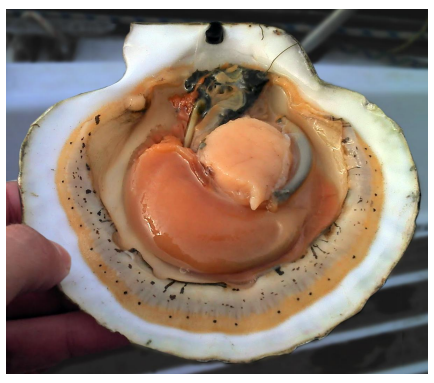 | oyster | shellfish | clam,<br>mollusk,<br>mollusc | Scallop,<br>bivalve |
| 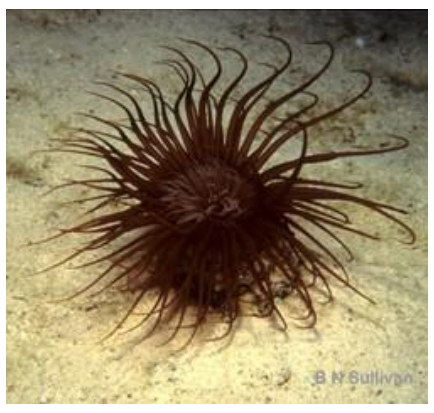 | sea urchin,<br>spiny starfish,<br>echinoidea | anemone | Tube anemone | Cerianthid |
| 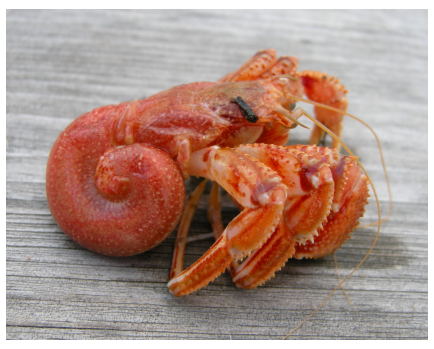 | shrimp,<br>lobster | crab | crustacean, | hermit crab,<br>pagurus |
| 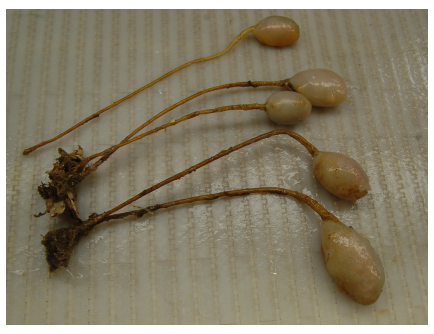 | kelp,<br>moss,<br>egg cases | seaweed,<br>plant | tunicate | stalked tunicate,<br>seaquirts |
| 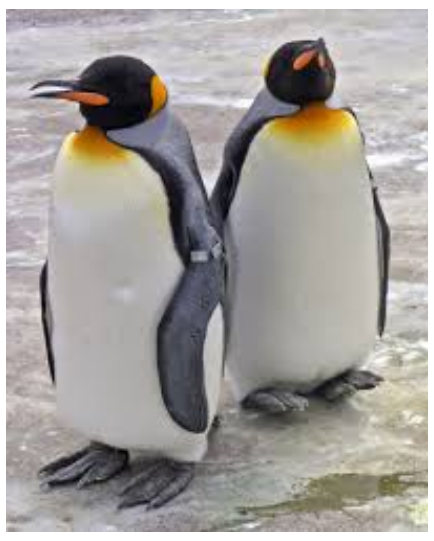 | Galapagos Island Penguin | Penguin,<br>Emperor | N/A | King,<br>Aptenodytes forsteri[10] |
| 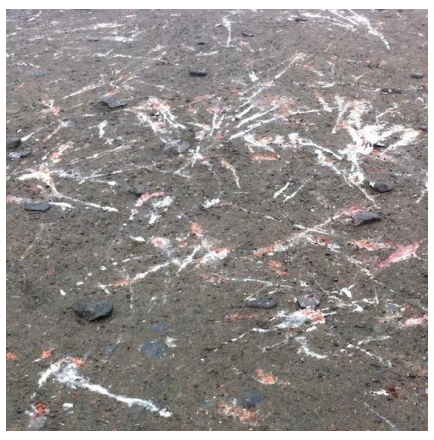 | Scuffling feet,<br>egg scraping | poo, shit, dung,<br>crap, dropping,<br>bodily waste | N/A | excrement,<br>guano,<br>fecal matter,<br>feces,<br>defecation,<br>evacuation<br>bowel<br>movement |
| 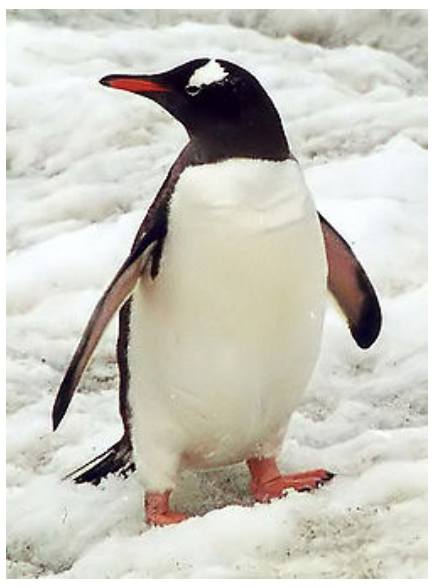 | Adelie, Jackass | Penguin | N/A | Gentoo,<br>Pygoscelis papua |

[10] Strictly this is an Emperor Penguin and therefore incorrect. But the knowledge of the term impressed us so we gave 3 points for this.

| | | | | |
|---|---|---|---|---|
| 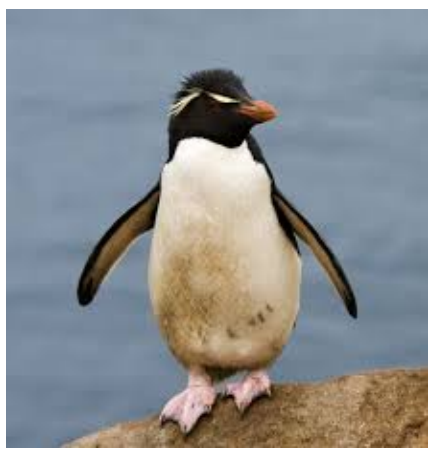 | Macaroni, Adelie | Penguin | N/A | Rockhopper |
| What is the gender of the above penguin? | Female or male. | Don't know, No idea, Unsure | N/A | Can't tell, Cannot tell/distinguish, DNA needed, blood test needed, not possible |

## Appendix B: Numeric Results

### 1. Descriptive Statistics

**Table B1: Descriptive Statistics of Variables Used in OLS.**

| Variable | Mean | Std Dev | Median | Min | Max |
|---|---|---|---|---|---|
| **Test performance** | | | | | |
| General Science Score | 7.04 | 3.16 | 7.00 | 0 | 15.00 |
| Project Specific Science Score | 6.76 | 3.44 | 7.00 | 0 | 15.00 |
| **Engagement Measures** | | | | | |
| Classifications | 249.10 | 1993.58 | 18.00 | 1.00 | 58,432.00 |
| Duration 1 (Days) | 639.82 | 403.33 | 633.29 | 21.45 | 156,433.00 |
| Duration 2 (Days) | 62.95 | 173.06 | 0.02 | 0 | 1235.30 |
| Active Days | 4.62 | 21.53 | 1.00 | 1.00 | 602.00 |
| **Project** | | | | | |
| Galaxy Zoo (Baseline) | 0.30 | 0.46 | - | 0 | 1.00 |
| Planet Hunters | 0.25 | 0.43 | - | 0 | 1.00 |
| Penguin Watch | 0.21 | 0.41 | - | 0 | 1.00 |
| Seafloor Explorer | 0.16 | 0.37 | - | 0 | 1.00 |
| Snapshot Serengeti | 0.09 | 0.28 | - | 0 | 1.00 |
| **Controls** | | | | | |
| Gender (female) | 0.44 | 0.50 | - | 0 | 1.00 |
| Age | 43.8 | 15.9 | 44.00 | 18.00 | 85.00 |
| Ethnicity (non-white) | 0.13 | 0.34 | - | 0 | 1.00 |
| Community Type (rural) | 0.34 | 0.47 | - | 0 | 1.00 |
| Education Level (9 pt scale) | 6.59 | 1.69 | 7.00 | 1.00 | 9.00 |
| Science Qualifications (yes) | 0.36 | 0.48 | - | 0 | 1.00 |
| Attitude to Science Learning in Zooniverse (Factor Score) | 0.00 | 1.00 | 0.11 | -5.53 | 2.08 |
| Opinion on Science Learning (Factor Score) | 0.00 | 0.99 | 0.11 | -3.98 | 2.10 |
| Religiosity (7 pt scale) | 2.83 | 2.12 | 2.00 | 0 | 7.00 |

## 2. Regression Statistics

**Table B2: OLS Regression Results: Dependent Variable = Ln(General Science Knowledge)**

| Variable | (i) Ln(Classifications) | (ii) Ln(Time since first classification) | (iii) Ln(Active period) | (iv) Ln(Active days) |
|---|---|---|---|---|
| Engagement | 0.011 | 0.014 | 0.001 | 0.010 |
| | | | | |
| Planet Hunters | -0.006 | -0.016 | 0.011 | -0.010 |
| Penguin Watch | -0.140** | -0.124* | -0.142** | -0.144* |
| Seafloor Explorer | -0.015 | -0.014 | -0.017 | -0.013 |
| Snapshot Serengeti | -0.096* | -0.088 | -0.089 | -0.089 |
| | | | | |
| Gender (Female) | -0.017 | -0.015 | -0.016 | -0.017 |
| Ln(Age) | 0.038 | 0.036 | 0.039 | 0.038 |
| Ethnicity (Non-White) | -0.009 | -0.009 | -0.015 | -0.009 |
| Community Type (Rural) | -0.009 | -0.009 | 0.013 | -0.009 |
| Education Level | -0.001 | -0.000 | -0.005 | -0.000 |
| Science Qualifications | -0.030 | -0.033 | -0.028 | -0.032 |
| Attitude to Science (Factor Score) | 0.095** | 0.095** | 0.102** | 0.095** |
| Opinion on Science Learning (Factor Score) | -0.024 | -0.023 | -0.014 | -0.024 |
| Religious Belief | 0.004 | 0.003 | 0.003 | 0.003 |
| Constant Term | 1.719** | 1.673** | 1.775** | 1.750** |
| | | | | |
| $R^2$ Value | 0.048 | 0.047 | 0.054 | 0.047 |
| F-Statistic | 6.820** | 6.666** | 7.114** | 6.685** |

**Note:** Statistical significance is denoted *=95% level; **=99% level

### Table B3: OLS Regression Results: Dependent Variable = Ln(Project Knowledge)

| Variable | (i) Ln(Classifications) | (ii) Ln(Time since first classification) | (iii) Ln(Active period) | (iv) Ln(Active days) |
|---|---|---|---|---|
| Engagement | 0.056** | -0.002 | 0.020** | 0.106** |
| | | | | |
| Planet Hunters | 0.286** | 0.265** | 0.275** | 0.259** |
| Penguin Watch | -0.284** | -0.310** | -0.262** | -0.296** |
| Seafloor Explorer | -0.211** | -0.198** | -0.204** | -0.208** |
| Snapshot Serengeti | -0.192** | -0.149** | -0.151** | -0.162** |
| | | | | |
| Gender (Female) | -0.046 | -0.042 | -0.030 | -0.048 |
| Ln(Age) | -0.009 | -0.010 | -0.008 | -0.017 |
| Ethnicity (Non-White) | 0.042 | 0.043 | 0.042 | 0.039 |
| Community Type (Rural) | 0.031 | 0.027 | 0.024 | 0.034 |
| Education Level | 0.003 | 0.005 | 0.004 | 0.004 |
| Science Qualifications | -0.022 | -0.031 | -0.018 | -0.027 |
| Attitude to Science (Factor Score) | 0.047** | 0.045** | 0.048** | 0.047** |
| Opinion on Science Learning (Factor Score) | 0.019 | 0.022* | 0.024* | 0.017 |
| Religious Belief | -0.001 | -0.003 | -0.001 | -0.001 |
| Constant Term | 1.680** | 1.869** | 1.870** | 1.827** |
| | | | | |
| $R^2$ Value | 0.155 | 0.129 | 0.143 | 0.155 |
| F-Statistic | 24.103** | 19.546** | 20.478** | 24.233** |

**Note:** Statistical significance is denoted *=95% level; **=99% level

**Table B4: OLS Regression Results Controlling for General Science Knowledge: Dependent Variable = Ln(Project Knowledge)**

| Variable | (i) Ln(Classifications) | (ii) Ln(Time since first classification) | (iii) Ln(Active period) | (iv) Ln(Active days) |
|---|---|---|---|---|
| Engagement | 0.052** | -0.006 | 0.020** | 0.104** |
| Ln(Science Knowledge) | 0.323** | 0.328** | 0.334** | 0.326** |
| | | | | |
| Planet Hunters | 0.306** | 0.289** | 0.288** | 0.281** |
| Penguin Watch | -0.242** | -0.271** | -0.217** | -0.252** |
| Seafloor Explorer | -0.201** | -0.189** | -0.193** | -0.199** |
| Snapshot Serengeti | -0.157** | -0.116** | -0.118** | -0.130** |
| | | | | |
| Gender (Female) | -0.038 | -0.034 | -0.025 | -0.040 |
| Ln(Age) | -0.016 | -0.016 | -0.012 | -0.024 |
| Ethnicity (Non-White) | 0.045 | 0.046 | 0.048 | 0.042 |
| Community Type (Rural) | 0.038 | 0.034 | 0.025 | 0.040 |
| Education Level | 0.005 | 0.006 | 0.006 | 0.005 |
| Science Qualifications | -0.018 | -0.025 | -0.017 | -0.023 |
| Attitude to Science (Factor Score) | 0.013 | 0.010 | 0.012 | 0.012 |
| Opinion on Science Learning (Factor Score) | 0.024 | 0.026* | 0.024 | 0.022 |
| Religious Belief | -0.001 | -0.004 | -0.002 | -0.002 |
| Constant Term | 1.095** | 1.283** | 1.234** | 1.223** |
| | | | | |
| $R^2$ Value | 0.235 | 0.214 | 0.228 | 0.239 |
| F-Statistic | 37.659** | 33.208** | 33.670** | 38.384** |

**Note:** Statistical significance is denoted *=95% level; **=99% level